\documentclass[]{article}

\usepackage[utf8]{inputenc}
\usepackage[a4paper, total={6.64in, 9.36in}]{geometry} 

\usepackage{hyperref} 
\usepackage{amsmath}
\usepackage{mathtools}
\usepackage{graphicx}
\usepackage{comment}
\usepackage{booktabs}
\usepackage{capt-of}
\usepackage{eurosym}
\usepackage{placeins}
\usepackage{authblk}
\usepackage{threeparttable}

\usepackage{pifont}
\newcommand{\cmark}{\ding{51}}%
\newcommand{\xmark}{\ding{55}}%

\providecommand{\keywords}[1]
{
  \small	
  \textbf{\textit{Keywords---}} #1
}

\usepackage{parskip}

\title{Three stages in the co-transformation of the energy and mobility sector}
\author[1,2,*]{Simon Morgenthaler}
\author[1,2]{Justus Dünzen}
\author[3]{Ingo Stadler}
\author[1,4]{Dirk Witthaut}

\affil[1]{Forschungszentrum J\"ulich, Institute for Energy and Climate Research, Systems Analysis and Technology Evaluation, 52428 J\"ulich, Germany}
\affil[2]{University Duisburg-Essen, Faculty of Engineering, Energy Technology, Lotharstr. 1, 47048 Duisburg, Germany}
\affil[3]{Technical University of Cologne, Cologne Institute for Renewable Energy, Betzdorfer Straße 2, 50679 Cologne, Germany}
\affil[4]{University of Cologne, Institute for Theoretical Physics, Z\"ulpicher Str. 77,
               50937 Cologne, Germany}
\affil[*]{Corresponding author: Simon Morgenthaler, s.morgenthaler@fz-juelich.de}

\date{\today}

\begin{document}
\maketitle

\noindent\makebox[\linewidth]{\rule{0.8\paperwidth}{0.4pt}}
\begin{abstract}
Renewable electricity sources such as wind and solar have shown a remarkable development in terms of efficiency, costs and availability, but system integration still remains a challenge.  Realizing a fully renewable electricity supply will require large scale storage technologies and flexible users to overcome long periods of low power generation. At the same time, other sectors such as mobility and industry must be electrified to replace fossil fuels. Power-to-Methane is a promising technology as it enables large-scale energy storage and sector coupling using existing infrastructures. In this work, we analyze the co-transformation of the German electricity, mobility and industry sector taking into account the recent decisions for coal phase out until 2038. We evaluate the necessary capacities of renewables and storage sizes as well as system costs and associated emissions using a techno-economic optimization model with a high technological and temporal resolution in the open source framework OSeMOSYS. We find three rather different stages of the transformation driven by the decreasing emission gap and the coal phase out. Solar power is expanded vastly until 2030, then coal is replaced mainly by fossil natural gas until 2040. Emission caps become very challenging afterwards such that all flexibility options are greatly expanded: Storage, curtailment and flexible Power-to-Methane. 
\end{abstract}

\noindent\makebox[\linewidth]{\rule{0.8\paperwidth}{0.4pt}}
\keywords{Synthetic natural gas, Power-to-Methane, Energy systems modeling, OSeMOSYS, Sector coupling, electromobility, residual load, capacity expansion}\\
\noindent\makebox[\linewidth]{\rule{0.8\paperwidth}{0.4pt}}

\section{Introduction}
\label{sec:introduction}

In order to move towards more sustainable development pathways in the electricity sector Germany has set ambitious targets to reduce its greenhouse gas emissions by phasing out of nuclear in 2022 and coal in 2038 \cite{klimaschutzplan2050}. The emerging generation gap will mostly be filled with intermittent renewable electricity generation, such as wind and photovoltaics. Hence, the integration of storage technologies is urgently needed to overcome long periods of missing power supply \cite{victoria2019sector}. 

At the same time, the mobility and industry sectors must be electrified to substitute fossil fuels. Battery electric vehicles (BEV) are rapidly entering the market for passenger traffic, but the decarbonization of air transportation and long distance freight transportation is far more challenging. These applications require energy carriers with a high specific energy such as hydrogen or synthetic hydrocarbons. The latter resources can also be a key for the decarbonization of the chemical industry, which is currently dependent on the usage of fossil natural gas and oil. Furthermore, many industry sectors need process heat at different temperatures, which is often provided by burning fossil fuels.

In this article we investigate the co-transformation of the electricity, mobility and industry sector in Germany until the year 2050. We focus on the role of two major technologies for sector coupling: (i) Electric mobility, i.e. the direct use of electricity in the mobility sector via BEVs, (ii) Power-to-Methane (PtM), i.e.~the production of synthetic natural gas (SNG) using renewable electricity for storage and sector coupling. The objective of this work is to investigate necessary capacities of renewables, storage operation and sizes that are needed for this transition of the German energy system as well as system costs and associated emissions.

PtM is a promising technology as it allows storing large amounts of renewable energy while using the existing gas grid infrastructure for distribution. Thus, SNG facilitates closing the discrepancy between supply and demand on a spatial and temporal level. SNG can be handled and utilized almost as natural gas such that no major technical adaptions are necessary. In particular, a reconversion to electricity is possible without additional CO$_2$ emissions within the existing gas fired power plants without modifications. Furthermore, PtM enables sector coupling as it can replace fossil natural gas (FNG) in a variety of applications, including in particular the chemical industry and potentially the production of synthetic fuels for the mobility sector.

To analyze the role of electromobility and PtM in the energy transformation, we set up an energy system model using the open source framework OSeMOSYS. Technologies are represented in an aggregated form with a high level of detail (more than 30 technologies are modeled, distinguishing between combined heat and power plants and different technological concepts, more details in section~\ref{sec:scenario-assumptions-and-model-input}). Each year consists of representative day types. A detailed time series analysis was conducted to find a suitable set of day types. It includes the temporal characteristics of electricity demand for both conventional grid demand and additional demand due to electric vehicles, wind speeds and solar radiation. Results show that the electricity sector is facing enormous challenges. Renewable capacities increase with growing demands of SNG and electricity for electric cars.

Against the background of future electricity demands, very high capacities of wind and photovoltaics are required for a resilient energy system. Therefore, a stronger European coordination in the energy sector could help to make optimal use of the different potentials of renewables. This is especially important since a higher share of renewables can also lead to critical conflicts of interest regarding ecological and social aspects, such as increasing land use requirements or social acceptance issues, for example. So, in order to carefully plan the future electricity system also non-technical aspects should be considered in order to account for possible negative side-effects on other sectors.

\section{Literature review}
\label{sec:literature}

The transformation of the energy system is necessary to limit global warming \cite{edenhofer2015climate}, however the particular pathways to stay below 1.5\,$^{\circ}$C or 2\,$^{\circ}$C are comparable \cite{rogelj2015energy}. Many studies find that shifting the energy system from fossil fuels to renewables is affordable \cite{plesmann_global_2014,jacobson2011providing,delucchi2011providing}. The multisectoral approach is very important, the focus must not only be on the electricity sector, although this sector is of particular importance \cite{Rockstrom2017}.

Within the electricity sector the share of renewable energy sources is rapidly increasing \cite{Edenhofer2011,Wiser2016}. Steadily decreasing costs for wind and PV favor this development, as both technologies are cost competitive \cite{irena2018renewable}. In particular PV capacity expansion is showing the most accelerated development and should not be underestimated \cite{creutzig2017underestimated}. The potential for global energy supply is very high, inter alia because of predictability \cite{haegel2019terawatt}. Victoria et al. find that on average one third of the electricity demand in Europe in a 95\,\% emission reduction scenario is covered by solar power \cite{victoria2019photo}. However, due to the intermittent generation character system integration remains a challenge. To successfully cope with non-dispatchable technologies the demand for flexibility increases \cite{elsner2015flexibilitatskonzepte}. Sector coupling is of special interest, as the approach provides flexibility and contributes to the decarbonization of other sectors \cite{creutzig2015transport, brown2018synergies, hoekstra2019underestimated, brown2019sectoral}.

In addition to the expansion of renewable energies, the transformation requires expanding energy storage facilities \cite{victoria2019sector}. Zsiborács et al. investigate the role of energy storage and emphasize their necessary capacity extension in the context of the European power system in 2040 \cite{zsiboracs_intermittent_2019}. While batteries are suitable for intra-day shifting and of high importance in PV dominated systems, gas storage systems are convenient for long term seasonal storage \cite{cebulla_how_2018, child_role_2018}. SNG can provide seasonal storage using existing gas grid infrastructure. Apart from reconverting SNG to electricity it can provide heat in the industry and building sector and utilized in the mobility sector as a transportation fuel.

The electrification of transport is a key instrument of the German federal government for greenhouse gas mitigation in the respective sector \cite{bundesregierung_klimaschutzprogramm}. Depending on charging and operation strategies integration of electromobility is beneficial for the system \cite{hanemann_effects_2017}. In particular, load shifting has a high potential \cite{BABROWSKI2014283}.  For applications which are harder to decarbonize, such as air transport, synthetic fuels are a promising option \cite{scheelhaase_synthetic_2019}.

Power-to-Methane is the technological basis to produce SNG. Researc\textbf{}h regarding this technology has increased steadily. This includes detailed research on the technological concepts \cite{ghaibPowertoMethaneStateoftheartReview2018} and economics \cite{gotzRenewablePowertoGasTechnological2016a}. Wulf et al. reviewed realized Power-to-Gas projects in Europe. A total of 128 projects were identified of which 36 produce synthetic natural gas, emphasising the importance of SNG \cite{wulfReviewPowertoGasProjects2018}. This is backed up by Thema et al., who identified 153 projects of which 60 produce SNG \cite{themaPowertoGasElectrolysisMethanation2019b}. Generally, costs for PtM are expected to decrease as more projects are realized and installed capacity increase \cite{schmidtFutureCostPerformance2017, schiebahnPowerGasTechnological2015}.

Former research using energy system modeling tools has identified Power-to-X technologies as a fundamental component of the German energy transformation process \cite{bauer_powerx_2020}. Within their multi-modal energy system planning approach for the case of Germany Müller et al. reveal that sector coupling facilitates an efficient transition \cite{muller_modeling_2019}. The long-term influence of electromobilty integration on the energy system is analyzed by Heinrichs \cite{heinrichs2014analyse}. A model suite developed by the Joint Research Center of the European Commission \cite{eu_science_hub} was applied by different research groups. Thiel et al. presented results of studies based on the electromobility modeling platform \cite{thiel_modelling_2016}. Parts of this group analyzed the impact of EU car policy regulations by applying an energy system model \cite{thiel_impact_2016}. Helgeson et al. focused on the European road transport and the electricity sector by means of a multisectoral partial-equilibrium investment and dispatch model \cite{HELGESON2020114365}. Open-source models with a high level of transparency have been developed as well. Examples are PyPSA-Eur \cite{horsch_pypsa-eur_2018}, Calliope Europe \cite{calliope_euro_trondle} and HTCOE \cite{MORGENTHALER2020114218}, which all provide the used data and thus enable a reproduction of results.

\section{Methods}
\begin{figure}[tb]
    \centering
    \includegraphics[width=0.85\textwidth]{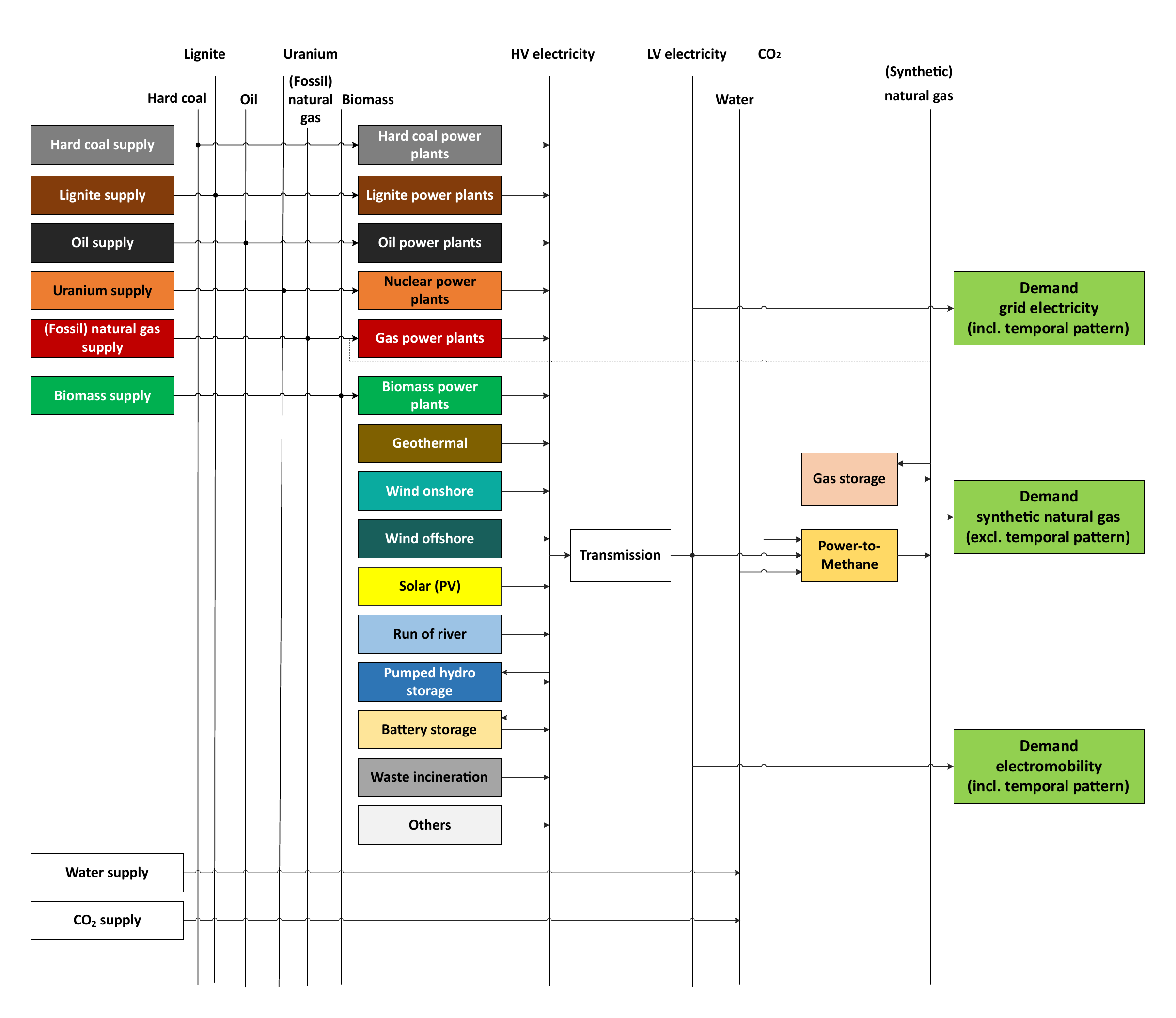}
    \caption{Reference energy system (RES). All vertical lines represent energy carriers which are considered in the model. Technologies, represented by boxes, can convert energy carriers into one another or import carriers to the system. The left side of the RES show supply technologies, the right side considered demand technologies. In between are conversion technologies.}
    \label{fig:ref}
\end{figure}

Our analysis is build on an energy system optimization model, which covers both operation and investments in the time period from 2018 to 2050. The model covers the German electricity sector including future sector coupling options for electromobility and PtM with a high technological and temporal resolution, but neglecting regional differences. Fig.~\ref{fig:ref} shows the basic set up of the model, where the boxes represent different technology classes. Vertical lines represent different energy carriers and horizontal lines show the in- and outputs of the technology classes. In the following we first introduce the structure of the model (section~\ref{sec:structure-optimization}) and then review the scenario assumptions and input parameters (section~\ref{sec:scenario-assumptions-and-model-input}). Finally, we describe how to achieve a high temporal resolution via time series aggregation (section~\ref{sec:temporal-patterns-and-resolution}).

\subsection{Structure of the optimization model}
\label{sec:structure-optimization}

The energy system optimization model is set up using the open source modeling system OSeMOSYS \cite{osemosys-web}. In the following we introduce the most important constituents of the model, while a complete mathematical description including all equations can be found online \cite{osemosys-code-long-math}.  We use the long code version including the storage equations (see \cite{osemosys-code-long} for more information about different code versions). 

\subsubsection{Objective function}

The objective function is minimizing the total discounted system costs for the entire model period from 2018 to 2050 (Eq.~\ref{eq:objective_function}). The objective function takes into account all regions (r), years (y) and technologies (t) and reads
\begin{equation}
\mathrm{
    min \sum_{y,r,t}{TotalDiscountedCost}_{y,r,t} \, .
    }
\label{eq:objective_function}
\end{equation}
An overview of used indices in all formula is provided in Table~\ref{tab:shortcuts_for_math}.

\begin{table}[tb]
    \centering
    \begin{threeparttable}
    \caption{Indexes for the sets for the mathematical description. The last column (n) states how many entries each set has in our model.}
    \label{tab:shortcuts_for_math}
    \begin{tabular}{lrr}
    \toprule
    Index & Set & n     \\
    \midrule
    t     & Technology & 39 \\
    r     & Region & 1      \\
    y     & Year & 33       \\
    e     & Emission & 1  \\
    l     & Timeslice & 288 \\
    f     & Fuel & 13 \\
    s     & Storage & 3 \\
    m     & Mode & 2\\
    \bottomrule
    \end{tabular}
    \end{threeparttable}
\end{table}

The total discounted system costs are calculated using the discounted investment and operating costs (fix and variable), possible discounted emission penalties and discounted salvage values for all technologies, regions and years (Eq.~\ref{eq:total_discounted_sys_costs}),
\begin{equation}
\begin{split}
    \mathrm{\forall_{y,r,t}{TotalDiscountedCost}_{y,r,t} = } \\
    \mathrm{DiscountedOperatingCost_{y,r,t}  } \\ +
    \mathrm{DiscountedCapitalInvestment_{y,r,t}  }\\ +
    \mathrm{DiscountedTechnologyEmissionPenalty_{y,r,t} } \\ -
    \mathrm{DiscountedSalvageValue_{y,r,t}} \, .
\end{split}
\label{eq:total_discounted_sys_costs}
\end{equation}
We assume a fixed discount rate of 5\,\%. Eq.~\ref{eq:discount_capital_investment} such that the discounted capital investments reads
\begin{equation}
\mathrm{
\forall_{y,r,t} \; {DiscountedCapitalInvestment}_{y,r,t} = \frac{CapitalInvestment_{y,r,t}}{(1 + DiscountRate_{r})^{y-StartYear}} \, .
}
\label{eq:discount_capital_investment}
\end{equation}
This method is used for all discounted parameters.

\subsubsection{Demand, production and use} 

The physical operation of the energy system and the technological characteristics of constituents are taken into account via a large set of equality and inequality constraints. In the following we review the most important constraints and its representation in OSeMOSYS. We start with the fundamental relations of demand, production and use.

Within each time step the production of all fuels must be greater or equal to the demand and use:
\begin{equation}
\mathrm{
    \forall_{y,l,f,r}{Production} \geq Demand_{y,l,f,r} + Use_{y,l,f,r} \, .
    }
    \label{eq:production_demand_use_timestep}
\end{equation}
The annual fuel production is the sum of the production in each time step:
\begin{equation}
\mathrm{
    \forall_{y,f,r}{ProductionAnnual}_{y,f,r} = \sum_{l}{Production}_{y,l,f,r} \, .
    }
\end{equation}
The condition from Eq.~\ref{eq:production_demand_use_timestep} is valid for the annual sums as well:
\begin{equation}
\mathrm{
    \forall_{y,f,r}{ProductionAnnual}_{y,f,r} \geq UseAnnual_{y,f,r} + AccumulatedAnnualDemand_{y,f,r} \, .
    }
\end{equation}
 
\subsubsection{Activity and Capacity} 

The maximum activity of a technology in each time step is tied to the available capacity:
\begin{equation}
    \begin{split}
    \mathrm{
    \forall_{y,t,l,r}{RateOfTotalActivity}_{y,l,t,r} \leq TotalCapacityAnnual_{y,t,r}} \times \\
    \mathrm{CapacityFactor_{y,l,t,r} \times CapacityToActivityUnit_{t,r}} \, .
    \end{split}
    \label{eq:ActivityCapacityLimit}
\end{equation}
The total capacity of a technology is the sum of new capacity (established during the modeling period as a result of the optimization process) and residual capacity (existing at the beginning of the modeling period as specified prior to the optimization process),
\begin{equation}
    \mathrm{
    \forall_{y,t,r}{TotalCapacityAnnual} = AccumulatedNewCapacity_{y,t,r} + ResidualCapacity_{y,t,r} \, .
    }
\end{equation}
The $\mathrm{CapacityFactor}$ describes the share of the available total capacity in each time step. This is particularly important for wind and solar power generation, where the availability depends on the weather. The units within the model must be consistent but can be freely selected. To link units for capacity (here in GW) and activity (here in TWh) to each other the parameter $\mathrm{CapacityToActivityUnit}$ is required. It describes the maximum activity that can be produced by one unit of capacity and has the dimension of time (here 8.76\,h).

\subsubsection{Storage operation}

As technologies, storages are represented using sets in OSeMOSYS. Each storage (which has a size in TWh) is linked to a technology (which has a capacity in GW) using the binary variables $\mathrm{TechnologyToStorage}$ and $\mathrm{TechnologyFromStorage}$. Charging and discharging of the storage is dependent on the activity in each time step of the technology, which is linked to the storage (Eq.~\ref{eq:storagecharge} and Eq.~\ref{eq:storagedischarge}).
\begin{equation}
    \begin{split}
    \mathrm{\forall_{y,t,r}{StorageCharge}_{s,y,l,r} = \sum_{t,m}{RateOfActivity}_{y,l,t,m,r} \times } \\ 
    \mathrm{TechnologyToStorage_{t,m,s,r} \times YearSplit_{y,l}}
    \end{split}
    \label{eq:storagecharge}
\end{equation}
\begin{equation}
    \begin{split}
    \mathrm{\forall_{y,t,r}{StorageDischarge}_{s,y,l,r} = \sum_{t,m}{RateOfActivity}_{y,l,t,m,r} \times } \\ 
    \mathrm{TechnologyFromStorage_{t,m,s,r} \times YearSplit_{y,l}}
    \end{split}
    \label{eq:storagedischarge}
\end{equation}
The variable $\mathrm{NetStorageCharge}$  is the difference between charging and discharging:
\begin{equation}
    \mathrm{
    \forall_{s,y,l,r}{NetStorageCharge_{s,y,l,r}} = StorageCharge_{s,y,l,r} - StorageDischarge_{s,y,l,r} \, .
    }
\end{equation}

\begin{table}[tb]
        \centering
        \begin{threeparttable}
        \captionof{table}{Most important input variables for the model, which have to be specified prior to the optimization process. The optimization variables are then determined by the optimization process which minimizes the discounted total system costs (Eq.~\ref{eq:objective_function}) over the entire model period. Dimensions are given in terms of the shorthands defined in Table \ref{tab:shortcuts_for_math}. A complete set of input variables and equations can be found in \cite{howells2011osemosys}.}
        \label{tab:important_inputs}
        \begin{tabular}{lrrr}
        \toprule
            Group & Variables & Unit & Dimensions\\
        \midrule
        Demands & AccumulatedAnnualDemand & TWh & r, f, y\\
              & SpecifiedAnnualDemand & TWh & r, f, y \\
              & SpecifiedDemandProfile &  & r, f, l, y\\
        
        \midrule
        Costs & CapitalCost &  Mio. \euro~/~GW & r, t, y\\
              & CapitalCostStorage &  Mio. \euro~/~TWh & r, s, y\\
              & FixedCost  & Mio. \euro~/~GW & r, t, y\\
              & VariableCost &  Mio. \euro~/~TWh & r, t, m, y\\
        \midrule
        Lifetimes & OperationalLife & Years & r, t \\
                  & OperationalLifeStorage & Years & r, s \\
        \midrule
        Existing infrastructure & ResidualCapacity & GW & r, t, y\\
                                & ResidualStorageCapacity & TWh & r, s, y \\
        \midrule
        Efficiencies & InputActivityRatio & & r, t, f, m, y \\
                     & OutputActivityRatio & & r, t, f, m, y \\
                     & CapacityFactor & & r, t, l, y \\
        \midrule
        Emissions & EmissionActivityRatio &  Mt / TWh & r, t, e, m, y\\
        \midrule
        Time representation & YearSplit & & l, y \\
        \midrule
        Constraints & AnnualEmissionLimit &  Mt & r, e, y\\
                  & TotalAnnualMaxCapacity & GW & r, t, y\\
                  & TotalTechnologyAnnualActivityUpperLimit & TWh & r, t , y \\
      \bottomrule
        \end{tabular}
        \end{threeparttable}
\end{table}

\subsubsection{Emissions}

The emissions of each technology is calculated using the parameter $\mathrm{EmissisonActivityRatio}$. It describes how much of an emission is emitted per unit of activity. In our model we focus on the aggregated GHG emissions measured in CO$_2$ equivalents, which are denoted as CO$_{2,eq}$. For the sake of simplicity we use the term CO$_2$ emissions as a synonym for CO$_{2,eq}$. For emissions which are dependent on the reaction stoichiometry of the combustion process (such as CO$_2$) it is reasonable to specify the parameter $\mathrm{EmissisonActivityRatio}$ for each technology that supply fuels to the energy system (cf. technologies on the left side in Fig.~\ref{fig:ref}). The total annual emissions of a technology are then given by
\begin{equation}
    \mathrm{
    \forall_{y,t,e,r}{AnnualTechnologyEmission}_{y,t,e,r} = \sum_{m}{AnnualTechnologyEmissionByMode}_{y,t,e,m,r} \, ,
    }
\end{equation}
where 
\begin{equation}
    \begin{split}
    \mathrm{
    \forall_{y,t,e,m,r}{AnnualTechnologyEmissionByMode}_{y,t,e,m,r} = } \\ 
    \mathrm{
    EmissionActivityRatio_{y,t,e,m,r} \times TotalAnnualTechnologyActivityByMode_{y,t,m,r} \, .
    }
    \end{split}
\end{equation}
To constrain the emissions a limit can be specified:
\begin{equation}
    \mathrm{
    \forall_{y,e,r}{AnnualEmissions}_{y,e,r} \leq AnnualEmissionLimit_{y,e,r},
    }
\end{equation}
where the quantity  $\mathrm{AnnualEmissionLimit}$ is an input parameter. We will discuss emission constraining in our model in more detail in section~\ref{sec:scenario-assumptions-and-model-input}.

\subsection{Scenario assumptions and model input}
\label{sec:scenario-assumptions-and-model-input}

The optimal transformation pathway of the energy system depends on a variety of factors which are external to the optimization and have to be specified beforehand to build the model. This includes in particular the demands for electricity (including electricity for electromobility) and synthetic natural gas as well as the emission cap. Additionally, techno-economic parameters of technologies as well as data about existing infrastructure must be specified. In this section we given an overview over the scenario assumptions and the model input used to set up the energy system model sketched in Fig.~\ref{fig:ref}. A summary of all important input variables are summarized in Table~\ref{tab:important_inputs}, which includes the variable names within OSeMOSYS as well the dimensions (sets) and the units we have used.  The most important techno-economic parameters for conversion technologies which we have used in this model are summarized in Table~\ref{tab:techno-economic-parameters}. We have aggregated the technologies to groups and only refer to these groups in our results for the sake of clarity.

\begin{figure}[tb]
    \centering
    \includegraphics[width=0.8\textwidth]{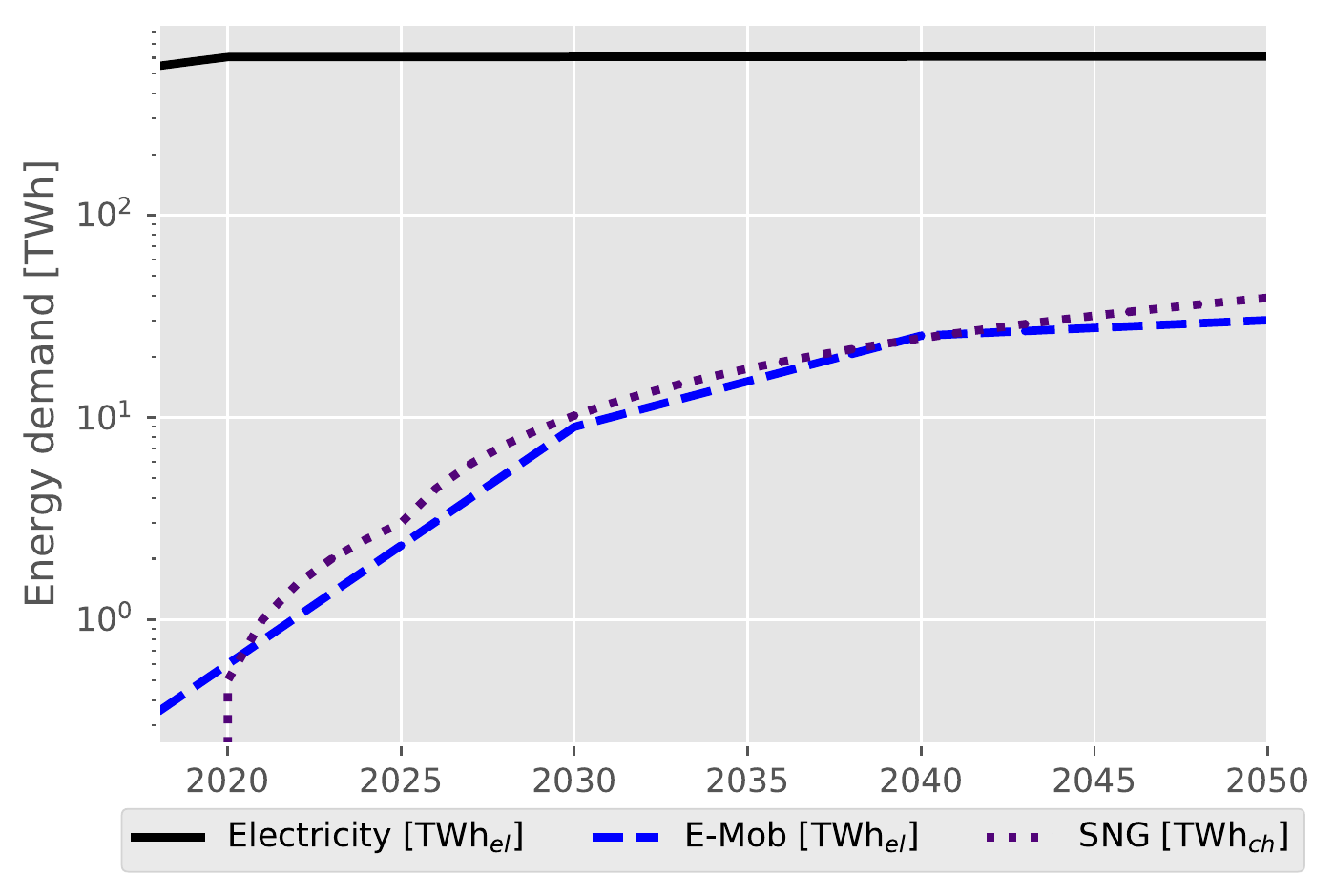}
    \caption{Demand for electricity, electricity for electromobility and SNG. The demand for electricity differ by their temporal pattern. The annual SNG demand can be supplied anytime during the year, allowing the PtM technology to provide flexibility to the system.}
    \label{fig:03demand}
\end{figure}

The main driver for the transformation of the energy system are the decreasing emission cap and the increasing demand for SNG and electricity for electromobility. In this article, we assume an emission reduction of 95\,\% until 2050 with respect to the reference year 1990 following the plan of the German federal government \cite{klimaschutzplanbmbu}. More precisely, we assume a linear decrease of the emission cap starting from 300.9\,Mt\,CO$_2$ in 2018 and decreasing to 21.35\,Mt\,CO$_2$ in 2050.

In the scenario which we present throughout this article we specify three fixed demands for electricity, electricity for electromobility and SNG (Fig.~\ref{fig:03demand}). The first two differ by their individual temporal pattern, which are described in more detail in section~\ref{sec:temporal-patterns-and-resolution}. They can be produced using the same electricity providing technologies. The electricity demand increases slightly at the beginning (in 2018 the demand is 545\,TWh$_{el}$) and then remains constant for the modeling period (from 2020 the demand is 605\,TWh$_{el}$). The demand for electricity for electromobility is very low at the beginning and increases exponentially until 2030. In the following decades we still assume an exponential increase for simplicity, albeit at lower growth rates, reaching a final value of 30\,TWh$_{el}$ in 2050. The demand for SNG starts in 2020 at a low value of 0.5\,TWh$_{ch}$ and increases to 39\,TWh$_{ch}$ in 2050. Demand is specified in steps of five years and is interpolated linearly in between. We note that these values only describe the final demand for SNG which is then distributed to other sectors or users outside the system boundaries. The total production of SNG can be higher, e.g.~for storage purposes, and is determined via the optimization.

\begin{table}[tb]
    \centering
    \begin{threeparttable}
    \caption{Techno-economic input parameters for our model. We distinguish between normal power plants (PP) and combined heat and power plants (CHP). Lifetimes, efficiencies and operation and maintenance costs are not object of change between the years. All parameters are taken from \cite{jesse2020potential} if not stated otherwise.}
    \label{tab:techno-economic-parameters}
	\begin{tabular}{lrrrrrrrr}
			\toprule
 			& Fuel Group & Renewable\tnote{1} & \multicolumn{3}{c}{Investment cost} &  O\&M cost & Lifetime  & Efficiency   \\

         	&         &  &  \multicolumn{3}{c}{[\euro$_{\rm year}$/kW$_{el}$]} &  [\euro/kW$_{el}$]  &       [a] &                         [\%]       \\
         	& &  &  2018 & 2030 & 2050 & & & \\
			\midrule
            CHP\_COA            &  Hard Coal & \xmark &           1646 &          1646 &          1646 &           33 &        45 &                          40         \\
            CHP\_LIG        &    Lignite & \xmark &            1872 &          1872 &          1872 &           40 &        45 &                          30          \\
            CHP\_NG-CCGT    &        Gas & (\xmark) &            823 &           823 &           823 &           21 &        40 &                          40          \\
            CHP\_NG-OCGT    &        Gas & (\xmark) &              568 &           568 &           568 &           17 &        40 &                          40          \\
            CHP\_NG-ST      &        Gas & (\xmark) &             1182 &          1182 &          1182 &           21 &        40 &                          31          \\
            CHP\_OIL        &        Oil & \xmark &            1182 &          1182 &          1182 &           20 &        40 &                          10          \\
            CHP\_OTH        &      Other & \xmark &            5500 &          5500 &          5500 &           55 &        20 &                           -          \\
            CHP\_WST        &      Waste & \xmark &            2530 &          1807 &          1807 &           51 &        30 &                           -          \\
            PP\_BIO         &    Biomass & \cmark &            2306 &          2210 &          2018 &          100 &        25 &                          28          \\
            PP\_COA         &  Hard Coal & \xmark &             1365 &          1365 &          1365 &           27 &        35 &                          38           \\
            PP\_GEO         &      Other & \xmark &             2400 &          2000 &          2000 &           84 &        30 &                           -          \\
            PP\_HYD-ROR     &      Hydro & \cmark &            1454 &          1575 &          1575 &           15 &       200 &                           -          \\
            PP\_LIG         &    Lignite & \xmark &             1856 &          1856 &          1856 &           39 &        35 &                          35          \\
            PP\_NG-CCGT     &        Gas & (\xmark) &             855 &           855 &           855 &           26 &        25 &                          58          \\
            PP\_NG-OCGT     &        Gas & (\xmark) &            568 &           568 &           568 &           17 &        40 &                          42          \\
            PP\_NG-ST       &        Gas & (\xmark) &             750 &           750 &           750 &           19 &        35 &                          42          \\
            PP\_NUC         &    Nuclear & \xmark &            5000 &          5000 &          5000 &           43 &        50 &                          33          \\
            PP\_OIL         &        Oil & \xmark &             750 &           750 &           750 &           20 &        40 &                          25          \\
            PP\_OTH         &      Other & \xmark &            5500 &          5500 &          5500 &           55 &        20 &                           -          \\
            PP\_SPV\tnote{2}&      Solar & \cmark &            460 &           384 &           236 &           20 &        30 &                           -          \\
            PP\_WOFF\tnote{3}&      Wind & \cmark &           2680 &          2380 &          1950 &          106 &        25 &                           -          \\
            PP\_WON\tnote{3}&      Wind & \cmark &           1296 &          1190 &          1110 &           34 &        25 &                           -          \\
            PP\_WST      &        Waste & \xmark &           2530 &          1807 &          1807 &           51 &        30 &                           -          \\
            PtM\tnote{4}       &      - &   -   &    2095 &           1498 &           503 &           15 &        15 &                          51          \\
            \bottomrule
	\end{tabular}
    \begin{tablenotes}
    \item[1] Classification used to calculate residual load in section~\ref{sec:residualload}. Gas is counted as renewable whenever SNG is burned. In the default mode of operation natural gas is burned and thus not counted as renewable.
    \item[2] Investment costs for 2018 based on \cite{irena-solar}.
            Cost degression assumptions derived from \cite{bartholdsen2019pathways}.
    \item[3] Based on \cite{jesse2020potential} (corrected to \euro$_{2018}$).
    \item[4] Based on \cite{THEMA2019775}. Lifetimes between PtM components, such as stack, balance of plant and methanisation reactor, may vary strongly. The chosen lifetime is a mixed value taking all components into account and based on own assumptions.
    \end{tablenotes}
    \end{threeparttable}
\end{table}

In addition to these global scenario assumptions, we have to specify a variety of input parameters for the respective technologies. Dispatchable conversion technologies need energy carriers (fuels) as an input (cf.~Fig.~\ref{fig:ref}). All energy carriers are made available by supply technologies (left side of the reference energy system, Fig.~\ref{fig:ref}). These energy carriers are associated with specific costs and CO$_2$ emissions, which are summarized in Table~\ref{tab:fuel-costs-and-emissions}. We assume constant costs for all energy carriers, while the investment costs for supply technologies may change. Cost and emission data for fuel supply is based on \cite{jesse2020potential}. Our model is calibrated to match the real electricity production shares in 2018 (see \cite{ise-electricitystatistics} for detailed data). However, we have corrected the specific emission factors to match the real CO$_2$ emissions (see \cite{uba-co2emissions} for detailed data) of the German electricity sector in 2018. 

The investment costs for solar photovoltaics are comparably small, reflecting the rapid technological development in recent years \cite{creutzig2017underestimated}. Earlier studies often assumed higher investment costs (e.g. \cite{jesse2020potential, morgenthalerSitedependentLevelizedCost2020c}), such that modeling output might differ significantly. However, the available areas for such installations are limited in the densly populated Germany. We set a capacity constraint to 200\,GW. This assumption is based on the projections of the international renewable energy agency \cite{internationalrenewableenergyagencyFutureSolarPhotovoltaic2019}. The same capacity constraint is set for wind onshore (200\,GW), however this figure is based ENSPRESO, a database for renewable potential in Europe \cite{ruizENSPRESOOpenEU282019a}.

\begin{table}[tb]
    \centering
    \begin{threeparttable}
    \caption{Fuel supply cost and emission data used in this model. Data is based on \cite{jesse2020potential}. Specific CO$_2$ emission factors are corrected to match the real emission data of the German electricity sector in 2018.}
    \label{tab:fuel-costs-and-emissions}
    \begin{tabular}{lrr}
    \toprule
         Fuel & Costs & Specific CO$_2$ emissions \\
              & [\euro/kWh] & [g/kWh$_{\mathrm{fuel}}$]\\
    \midrule
        Hard coal & 0.0061 & 0.412 \\
        Lignite & 0.0013 & 0.42 \\
        Oil & 0.026 & 0.28 \\
        Natural gas & 0.0207 & 0.314 \\
        Uranium & 0.0033 & 0\\
        Biomass & 0.001 & 0 \\
   \bottomrule
    \end{tabular}
    \end{threeparttable}
\end{table}

We consider three types of storage (cf.~Fig.~\ref{fig:ref}): battery, pumped hydro and the natural gas grid. Table~\ref{tab:storage-techno-economic-parameters} gives an overview of the used techno-economic parameters for storage technologies. Values for efficiencies are for a round-trip. The capacity of the  gas grid is given in chemical energy (TWh$_{ch}$) of SNG, while the capacity of hydro and battery storage is given in electrical energy (TWh$_{el}$).

The residual capacity given in Table~\ref{tab:storage-techno-economic-parameters} is the existing capacity in Germany's energy system in 2018. Capacity expansion is allowed for battery storage only. An expansion of pumped hydro storage is facing severe limitation due to the lack of suitable locations and the lack of public acceptance. The capacity of the gas grid is huge already today.

\begin{table}[tb]
    \centering
    \begin{threeparttable}
    \caption{Techno-economic input data for storage. We assume no cost degression. Capacity expansion is only allowed for the battery storage.}
     \label{tab:storage-techno-economic-parameters}
    \begin{tabular}{lrrrrrr}
    \toprule
    & Investment & O\&M & Efficiency & Lifetime & Residual & Expansion \\
    & cost & cost & [\%] & [yrs] &  capacity & possible \\
    \midrule
    Battery\tnote{1} & 109 \euro/kWh & 6.4 \euro/kW & 95.4 & 15 & 300 MWh$_{el}$ & \cmark \\ 
            & 140 \euro/kW  & 1.9 \euro/MWh &     &    & &       \\
    Pumped hydro\tnote{2} & - & 28 \euro/kW & 75 & 200 & 38.5 GWh$_{el}$ & \xmark \\
    Gas grid\tnote{3} & - & - & 100 & 100 & 231 TWh$_{ch}$ & \xmark \\
    \bottomrule
    \end{tabular}
    \begin{tablenotes}
    \item[1] Based on \cite{pfenninger2017ukmodel, pfenningerModelUKPower}.
    \item[2] Based on \cite{jesse2020potential, pohlerPumpspeicherKraftwerkePSWBayerischer, stenzelEnergiespeicherJahresubersicht2017}.
    \item[3] Based on \cite{statistaKapazitatUntertageErdgasspeicherDeutschland2020, stronzikWettbewerbImMarkt2008}.
    \end{tablenotes}
    
    \end{threeparttable}
    
\end{table}

\begin{figure}[tb]
    \centering
    \includegraphics[width=0.8\textwidth]{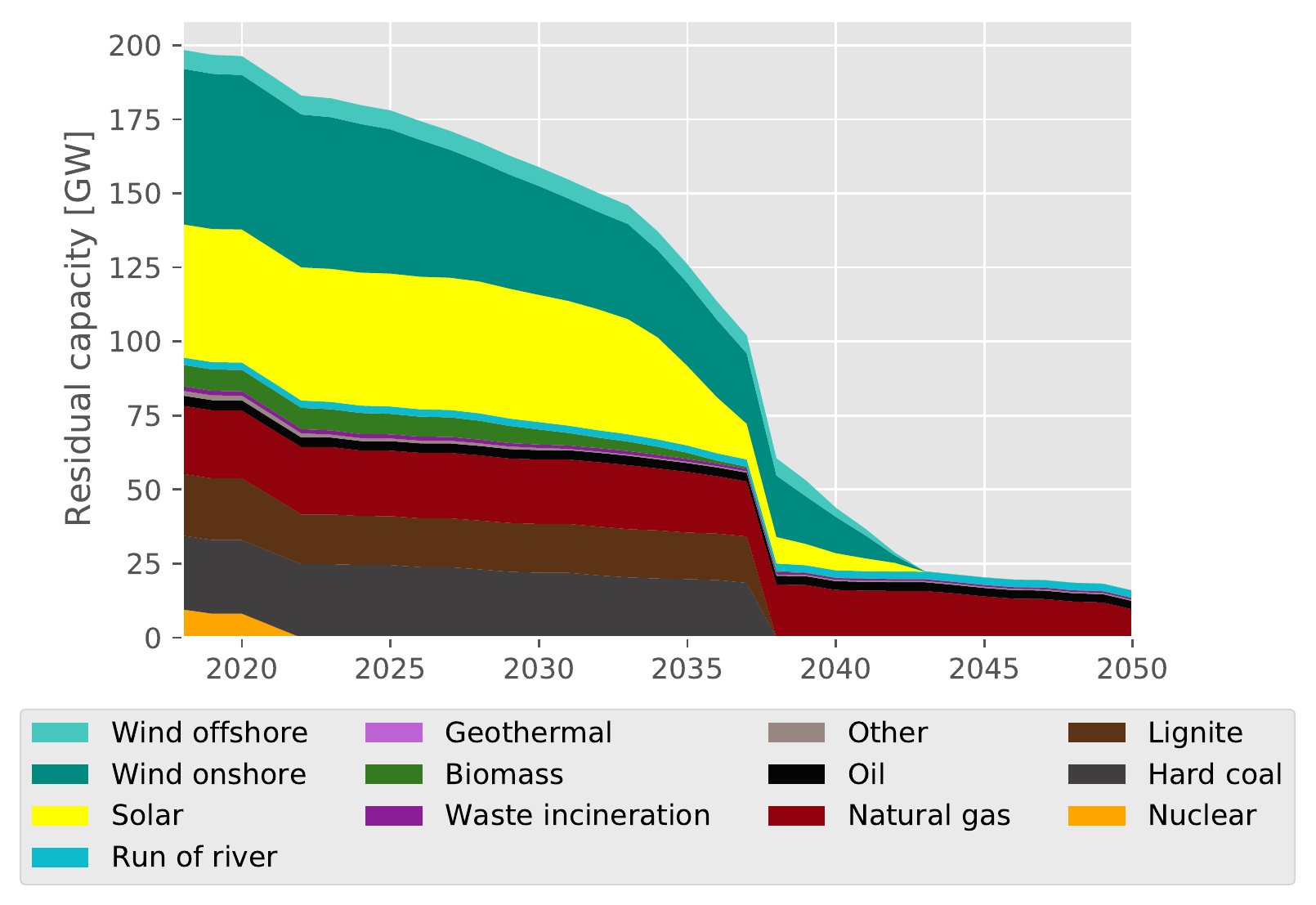}
    \caption{Existing energy infrastructure in Germany which we use as residual capacity. Residual capacity is already available at the beginning of the modeling. The aging structure is a result of our analysis using the open source Python package powerplantmatching \cite{gotzens_performing_2019, ppm}. For the existing infrastructure no investment costs are considered, if capacities are utilized associated costs are O\&M (cf.~Table~\ref{tab:techno-economic-parameters}) and (if applicable) fuel costs (cf.~Table~\ref{tab:fuel-costs-and-emissions}) only.}
    \label{fig:04residualcapacity}
\end{figure}

We have used the open source Python package powerplantmatching to identify existing energy infrastructure in Germany \cite{gotzens_performing_2019}. The code is freely available on Github \cite{ppm}. This package provides detailed information about existing capacities of power plants including their commissioning and possibly retrofit date. The data is compiled using different power plant databases, standardizing data formats and merging data. Using this information capacities for power plant types (as presented in Table~\ref{tab:techno-economic-parameters}) can be aggregated. Based on the commissioning date of each power plant and its lifetime we have calculated the expected decommissioning date. We used this information to calculate decommissioning curves (Fig.~\ref{fig:04residualcapacity}). That is the residual capacity which is available to the model. Using this capacity is only associated with operation and maintenance costs as well as fuel costs (where applicable). Coal power plants and nuclear power plants are an exception, as the phase out is politically decided. Nuclear phase out will be in 2022, coal phase out at the latest in 2038.

It should be noted that we don't consider import and export options (cf.~Fig.~\ref{fig:ref}), which can potentially create synergies in an interconnected energy system. However, a resilient energy system should be designed to operate self-sufficiently. In addition, import would typically appear whenever renewable generation is low, but simultaneously electricity generation in neighbouring countries is probably low as well.

\subsection{Temporal patterns and resolution}
\label{sec:temporal-patterns-and-resolution}

To capture the variability of renewable power generation and consumption a high temporal resolution of the optimization model is necessary. The variability of renewable power generation is modeled using historic data, using bias-corrected capacity factors of wind and solar power generation in Germany with an hourly resolution from renewables.ninja \cite{pfenninger_renewables_1, staffell_renewables_2, renewables_3}. We use capacity factors from 2012 as former research has shown that this year is a representative single weather year for renewable power generation \cite{collins2018representative}. The actual variable renewable power generation in the model is obtained by multiplying the capacity factor (model input) with the installed generation capacity, which is subject to the optimization. 

The demand for electricity is a model input and is also provided at an hourly resolution. We use an electricity load curve for Germany in 2016 from the open power system database \cite{opsd_loadcurve_data}. Characteristics of electromobility load curves are heavily dependent on user behaviour and the future integration of smart charge management systems. For this model we distinguish between weekday and weekend simultaneity profiles only, based on an analysis from \cite{linssen2019modellierung}. We do not include an optimization of the charging process here, as the dispersion and actual usage of this technology is unforeseen. 

\begin{figure}[tb]
    \centering
    \includegraphics[width=0.65\textwidth]{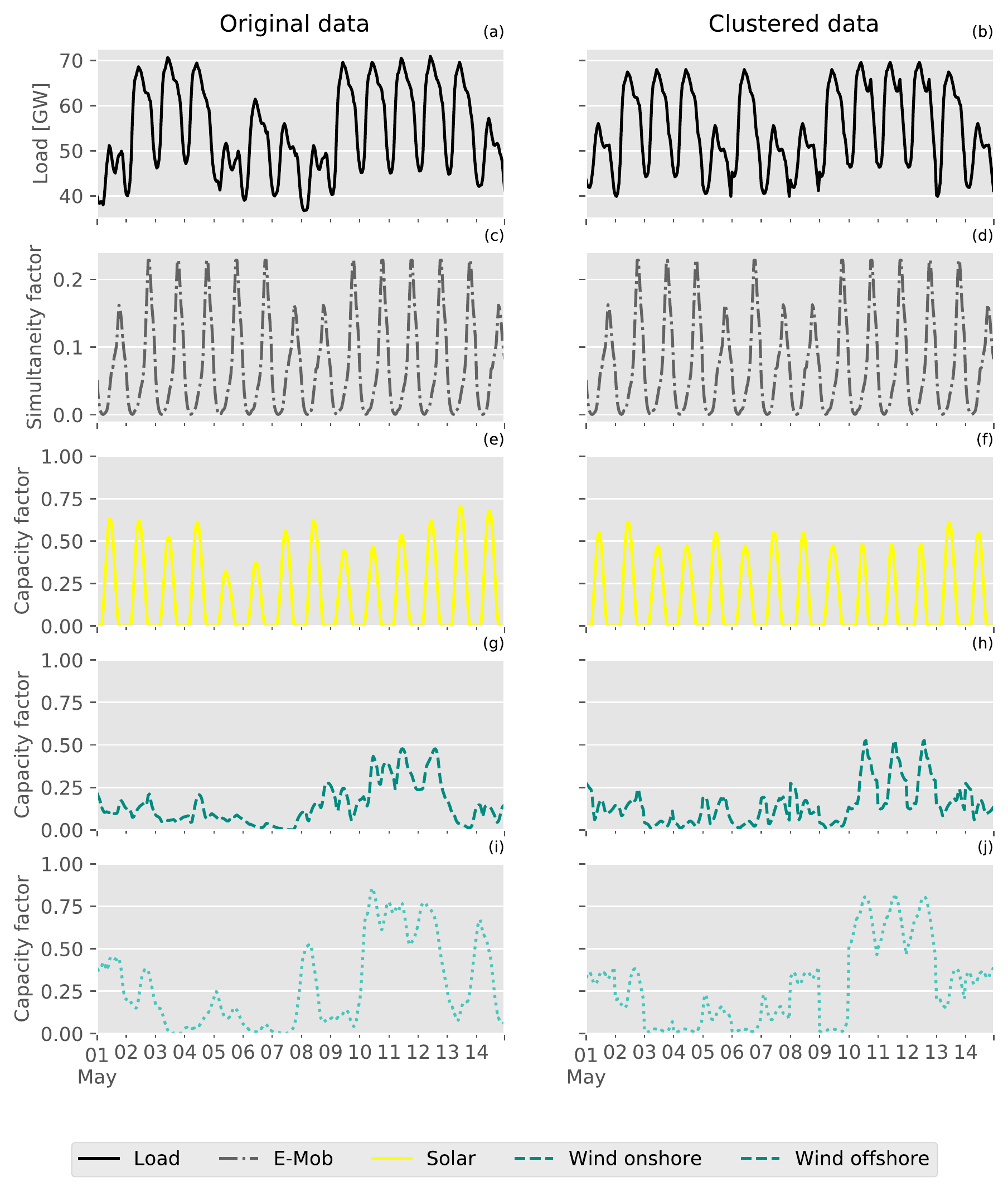}
    \caption{Original and clustered data for two weeks in May. Shown are the electricity load, the simultaneity charging profile for electromobility and the capacity factors for renewable energy technologies. The clustering is a result of our analysis with TSAM \cite{KOTZUR2018474}. The annual original time series have an hourly resolution , we obtain twelve typical days with an hourly resolution from our analysis with TSAM. Clustering is necessary to deal with the computational burden. This allows the optimization process to take place within a reasonable time.}
    \label{fig:original_vs_clustered}
\end{figure}

An example of renewable power generation (capacity factors), electricity load curves and the simultaneity factor for electromobility for two weeks in May is shown in Fig.~\ref{fig:original_vs_clustered} (left column, original data). 

A numerical optimization with an hourly resolution for more than 30 years represents an extraordinary computational burden, thus a simplification of the problem is needed. Reducing the temporal resolution is not practical, as the variability of renewable generation cannot be captured adequately. However, time series data may have repetitive characteristics. For instance,  electricity load profiles are similar every workday and solar power has a characteristic daily profile with a peak around noon differing only in magnitude from day to day. These characteristic pattern can be used to define typical time periods, which capture the essential aspects of intra-day and inter-day variability. 

\begin{figure}[tb]
            \centering
            \includegraphics[width=\columnwidth]{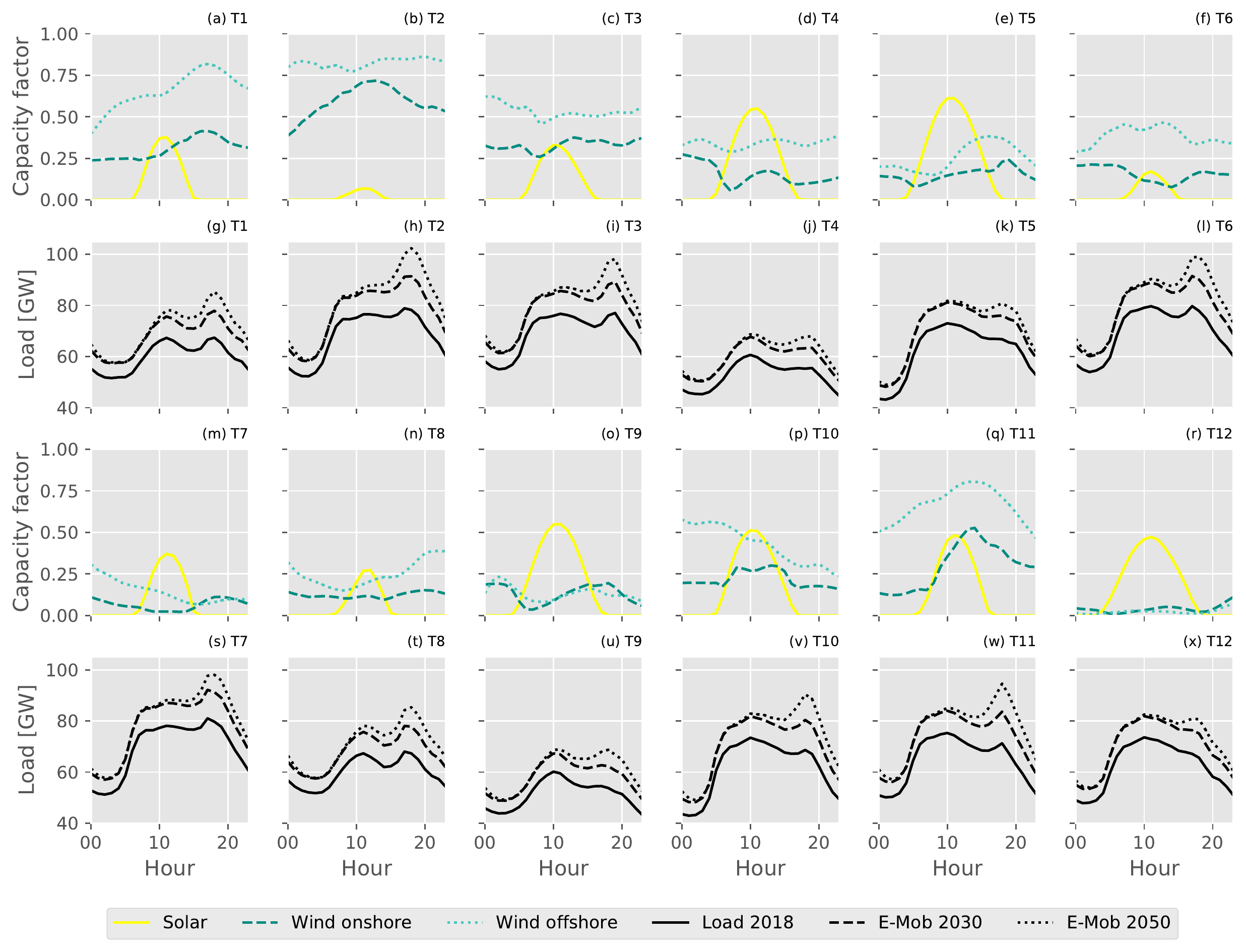}
            \caption{Characteristics of the twelve type days. Shown are the capacity factors for renewable energy technologies and the electricity load for 2018, 2030 and 2050. The electricity load is a result of electricity demand and additional demand from electromobility. The annual electricity demand is shown in Fig.~\ref{fig:03demand}. As the SNG demand can be supplied whenever it is cost efficient the resulting electricity for PtM facilities is not part of the load characteristics.}
            \label{fig:typedays4x6}
        \end{figure}

The open-source python package time series aggregation module (TSAM) is capable of clustering multi-dimensional time series data \cite{KOTZUR2018474}. We have used this module to cluster the time series data for solar, wind onshore, wind offshore, electricity load and electromobility load to create twelve typical periods consisting of 24 hours each (type days). Thus, the amount of time steps per year, which have to represented explicitly in the optimization model, is reduced from 8760 to 288 (that is approx. 3.3\,\% of the original data). The type days are weighted according to the frequency of occurrence in original data, which is also provided by TSAM (cf.~Table~\ref{tab:typedays_weight}). The clustered type days are shown in Fig.~\ref{fig:typedays4x6}. The total electricity load is a function of the demand for electromobility electricity, which increases during the model period. This is shown with the three different black lines (base load for 2018 without electromobility and resulting electricity load shifting due to electromobility in 2030 and 2050). 

\begin{table}[tb]
    \centering
    \begin{threeparttable}
    \caption{Weighting of the twelve type days in our model. The weighting is a result of TSAM and is based on the occurrence frequency in the original data.}
    \label{tab:typedays_weight}
    \begin{tabular}{lrr}
    \toprule
     & Days per year & Share [\%] \\
     \midrule
     T1 & 23 & 6.3 \\
     T2 & 22 & 6 \\
     T3 & 27 & 7.4 \\
     T4 & 38 & 10.4 \\
     T5 & 52 & 14.2 \\
     T6 & 27 & 7.4 \\
     T7 & 36 & 9.9 \\
     T8 & 23 & 6.3 \\
     T9 & 27 & 7.4 \\
     T10 & 24 & 6.6 \\
     T11 & 22 & 6 \\
     T12 & 44 & 12.1 \\
     Sum & \textbf{365} & \textbf{100} \\
   \bottomrule
    \end{tabular}
    \end{threeparttable}
\end{table}

A comparison of the clustered data with the original data for the first two weeks of May (cf.~Fig.~\ref{fig:original_vs_clustered}) reveals that accuracy differs. For example, the daily peaks in the original electricity load curves are slightly flatter in the clustered curve. The 5th of May in 2016 was a holiday, thus the peak has a weekend profile. This is correctly depicted in the clustered time series. However, the following day (a Friday) is an intermediate day between holiday and weekend. As many people take the day off the load is a bit lower than during the week, but not as low as on the weekend. This detail cannot be reproduced in the clustered time series. A possible reason is that not many equal days exists, such that it is not considered in the clustered data. A comparison of clustered and original data for capacity factors show that the essential intra-day and inter-day variability of renewable power generation is captured, but obviously not all details of the time series can be reproduced.

The demand for Power-to-X (PtX) is handled differently than the demand for electricity. In this article, we have settled on SNG as a representative PtX fuel as it offers numerous advantages discussed in detail in section~\ref{sec:introduction}.  In particular, SNG can be easily stored in large amounts in the existing gas grid infrastructure (we assume a storage size of more than 200\,TWh$_{ch}$ in Germany, cf.~Table~\ref{tab:storage-techno-economic-parameters}). Hence, it is sufficient to specify the the accumulated annual demand for SNG and we neglect its temporal variability.

The actual production of SNG per hour and the capacity of PtM is subject to the optimization. This allows to produce SNG whenever it is most cost effective in terms of total system costs (cf.~section~\ref{sec:structure-optimization}, Eq.~\ref{eq:objective_function}). Therefore, the option to build more PtM capacity than would be necessary is available ("excess capacity"). In this case PtM may act as a flexible consumer, providing flexibility to the system through demand side management. In particular, this enables a forward shifting of the production to satisfy later demands over longer periods of time. 

Additionally, we included a second mode of operation for all considered gas fired power plants. In this mode, SNG instead of FNG is burned for electricity generation. Efficiencies and operation and maintenance costs are equal for both modes. However, SNG has a neutral CO$_2$ footprint. It should be noted that the round-trip efficiency from electricity to electricity is $\mathrm{\eta_{PtM} \times \eta_{PP_{NG-CCGT}} = 30\,\%}$ in the best case. For the sake of simplicity we neglect possible ramping times and minimum loads of the PtM process.


\section{Results}

\subsection{Three stages of the transformation}
\label{sec:stages-of-transformation}

\begin{figure}[tb]
    \centering
    \includegraphics[width=0.8\textwidth]{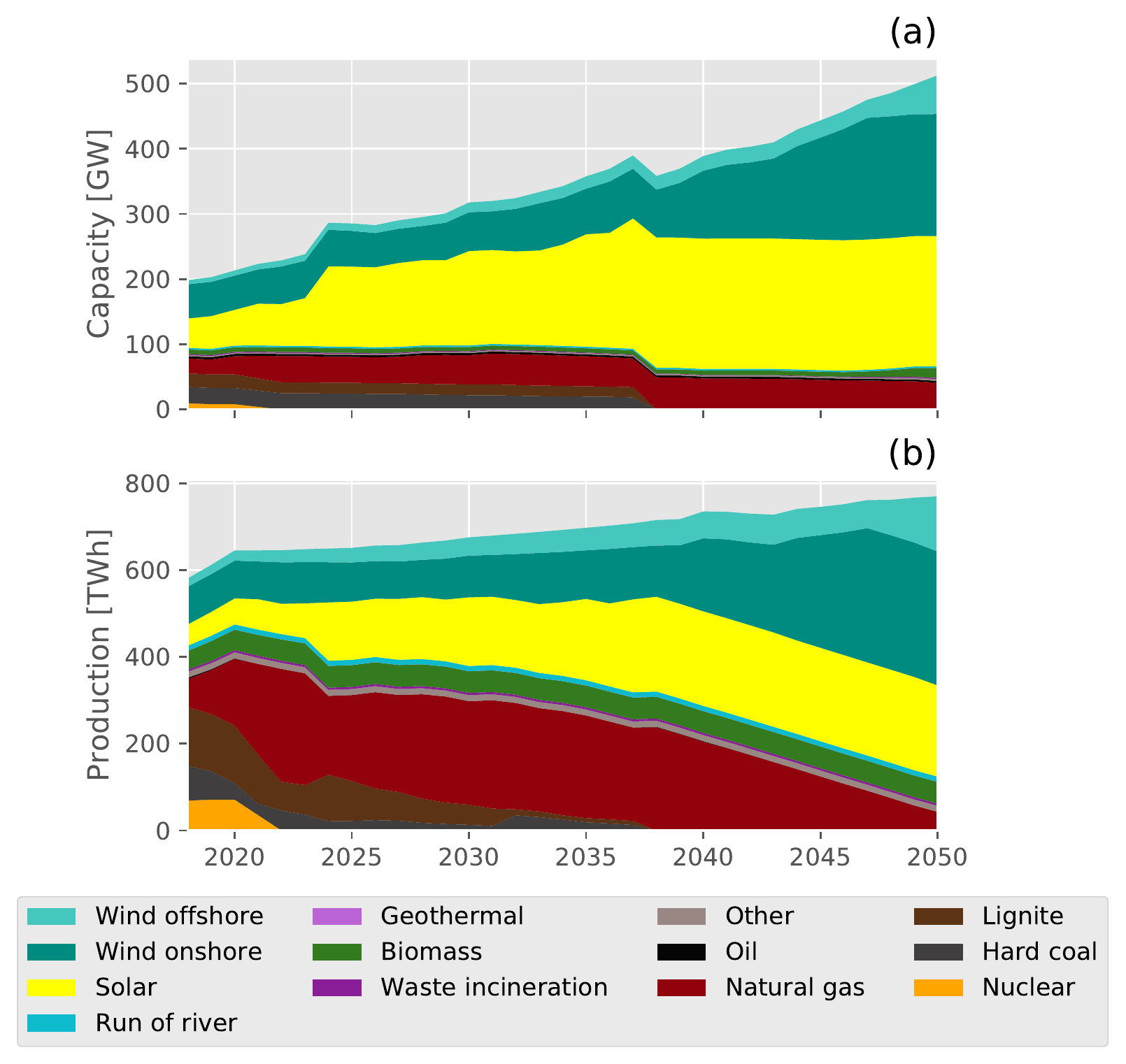}
    \caption{(a) Capacity mix development from base year 2018 to 2050. The capacity mix for the base year is set to the capacity mix of the German energy system. The development is a result of the optimization process. It shows different stages of the transformation process, shifting from fossil fuels to renewables. Nuclear phases out in 2022, coal in 2038. (b) Plots the actual electricity production mix. Average capacity factors of renewables are significantly smaller than those of dispatchable technologies. The maximum capacity of photovoltaics and wind onshore is constrained to 200\,GW (based on \cite{internationalrenewableenergyagencyFutureSolarPhotovoltaic2019, ruizENSPRESOOpenEU282019a}). Other capacities are constrained to their potential (hydro, biomass).
    }
    \label{fig:capacity_expansion}
\end{figure}

The optimum co-transformation pathway consists of three stages with strongly different characteristics: First, renewable capacities are rapidly increased before coal is phased-out and replaced mainly by natural gas. In the final decade, fossil fuels are phased out almost completely, restricting the options for conventional backup generation, such that all modes of flexibility are needed. These three stages are described in detail in the following.

\subsubsection{2018-2030: Rise of photovoltaics}

The first decade is dominated by a rapid expansion of renewable generation capacities (Fig.~\ref{fig:capacity_expansion}). Photovoltaics is preferred for expansion, as the specific investment costs are significantly lower than those of other technologies, in particular wind power (cf.~Table~\ref{tab:techno-economic-parameters}). The capacity of solar PV is almost tripled, from 55\,GW in the base year 2018 to 145\,GW in 2030. The same applies to the electricity production (capacity is fully utilized, curtailment of electricity from photovoltaics is close to zero (cf.~Fig.~\ref{fig:curtailment})). 

\begin{figure}[tb]
    \centering
    \includegraphics[width=0.8\textwidth]{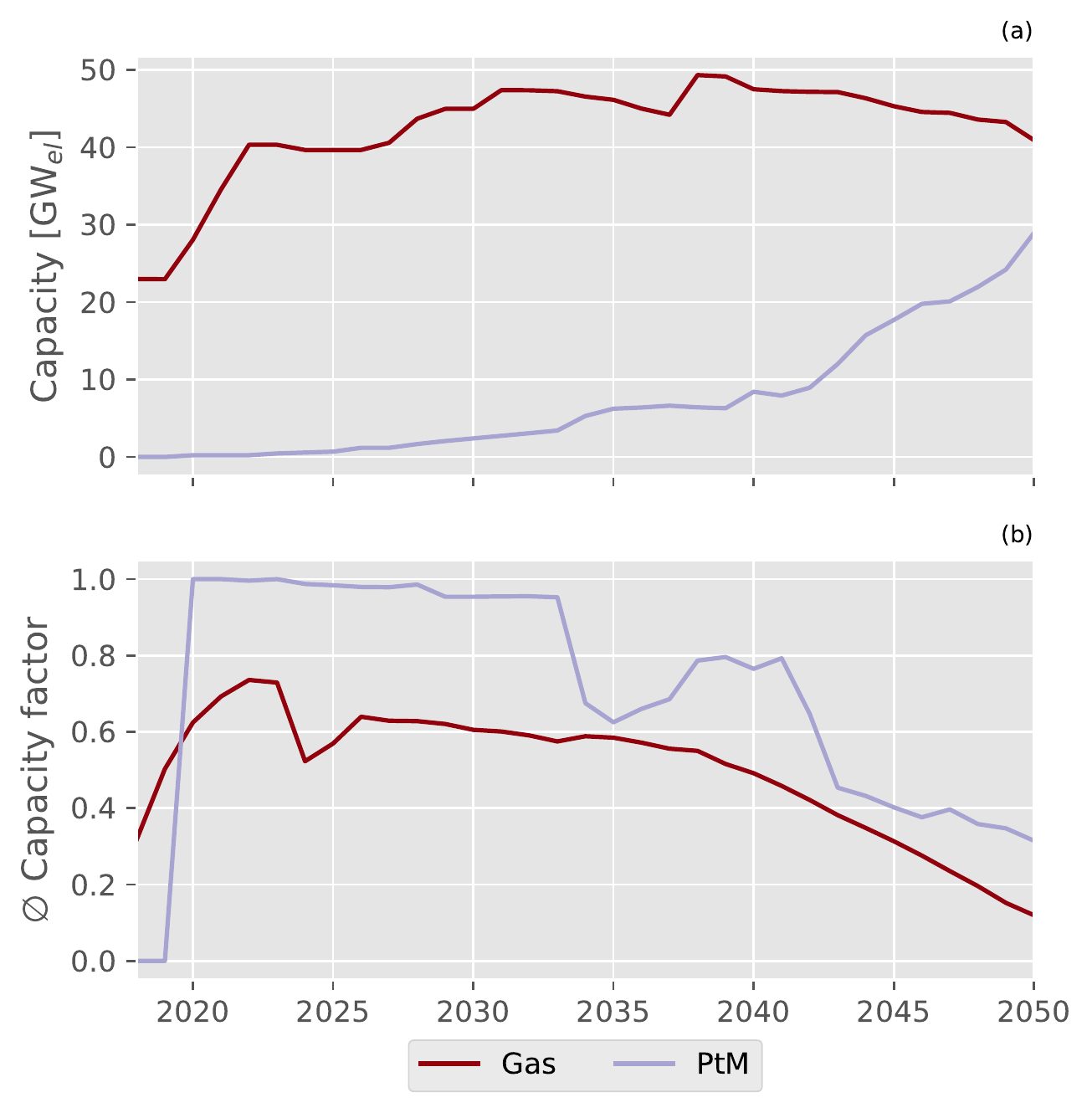}
    \caption{(a) Capacity and (b) capacity factor development of gas fired power plants and PtM. Both technologies are crucial for providing system flexibility. Their task change over the modeling period, which is discussed in detail in section~\ref{sec:ptgop}~\&~\ref{sec:residualload}.}
    \label{fig:cap_vs_fng_ptm}
\end{figure}

As nuclear capacity drops out by 2022 its production of 70\,TWh must be replaced while complying with increasingly stringent emission limits. Photovoltaics can provide this energy only when weather conditions are favorable, such that natural gas is needed for backup. As a consequence, both the capacity and the utilization of gas fired power plants increases strongly. The utilization rate jumps from 35\,\% in 2018 to a peak value of 75\,\% in 2022 directly after the nuclear phase out and settles at around 60\,\% in 2030 (cf.~Fig.~\ref{fig:cap_vs_fng_ptm}). 

The gas fired power plants can readily adapt to the load, which is sufficient to provide the necessary flexibility for the electricity system. Storage plays no significant role during this stage of the transformation. Only the residual battery storage with a capacity of of 300\,MWh (cf.~Table~\ref{tab:storage-techno-economic-parameters}) is used and no storage capacity expansion takes place.

During the first decade the SNG demand is quite low (cf.~Fig.~\ref{fig:03demand}). Due to the low demand the installed capacity in 2030 of PtM is 2.4~GW$_{el}$, with a high utilization of above 95\,\%. Thus, it does not contribute to any flexibility options yet, as it runs almost continuously.

\subsubsection{2030-2040: Coal-phase out}

The decade from 2030-2040 is distinguished by the phase-out of coal power plants. According to the plans of the German federal government, the last coal-fired power plant is taken out of operation in 2037 \cite{kohleausstieg_komission}. The missing energy is mainly provided by fossil natural gas (FNG), capacities are almost doubled reaching a peak in 2038 with 50\,GW while average capacity factors remain high (above 50\,\% until 2040, cf.~Fig.~\ref{fig:cap_vs_fng_ptm}). Replacement of coal by gas is sufficient to satisfy the emission reduction targets for the most part. Hence, the expansion of renewable energy sources is slowed down considerably in this decade. 

Strong changes are observed in terms of system operation. FNG plants remain the most important provider of flexibility, but can no longer handle everything. Other technological options are needed which provide flexibility. Renewable curtailment is used as a flexibility option in the early 2030s, but reduces after the final coal phase-out (cf.~Fig.~\ref{fig:curtailment}). After coal phase-out more renewable energy is needed such that renewable curtailment can no longer offer the needed flexibility. Instead, batteries now take an important role. The storage capacity of batteries start to increase from 2030 with a steep increase just before the final coal phase-out in 2038. Its capacity steadily increase afterwards. The charging and discharging cycles of battery storage make up to almost 50\,TWh per year from 2038 onwards. Batteries are mostly used for daily shifting of electricity from photovoltaic peaks to satisfy the increasing peak demand due to electromobility in the evening (cf.~Fig.~\ref{fig:typedays4x6}). 

The average capacity factor of PtM remains above 95\,\% until 2034. From this year onwards the PtM technology provide flexibility increasingly, with an average capacity utilization of approximately 75\,\% (Fig.~\ref{fig:cap_vs_fng_ptm}). This means that more capacity is build than would be required to meet the SNG demand (excess capacity). PtM decreases SNG production during the daily evening peaks and during times of low renewable generation (cf.~Fig.~\ref{fig:ptm_cfs_typedays}). Thus, PtM takes over an important role within the system at an early stage, this role will become even more important in the final decade of the modeling period.

\subsubsection{2040-2050: Full flexibility}

Emission reduction targets become challenging in the decade 2040-2050 causing a massive change in energy system operation (cf.~section~\ref{sec:scenario-assumptions-and-model-input} for emission limits). Previously, FNG was replacing coal due to the lower specific emissions -- but this is no longer enough. Now also FNG power generation must be replaced by zero-emission technologies. Remarkably, FNG capacities decrease only slowly, but the operation changes fundamentally. From 2040 the average capacity factor of gas fired power plants decrease from 50\,\% to below 15\,\% in 2050, while capacities only drop by 5\,GW in this period (cf.~Fig.~\ref{fig:cap_vs_fng_ptm}). Partly, this can be explained by the lifetime of these technologies (cf.~Table~\ref{tab:techno-economic-parameters}), as they were required earlier in the modeling they still exist. However, large parts of the gas capacities are still required for backup purposes. We discuss the development of residual load change in detail in section~\ref{sec:residualload}.

As power generation from FNG is reduced, renewables have to stand in. As a consequence we see a further rapid increase of solar photovoltaic capacities until the capacity constraint of 200\,GW is met. Afterwards, wind power generation is expanded. In particular, the capacity of onshore approximately doubles from 2040 to 2050 while offshore wind power capacity triples. However, the capacity constraint for wind onshore of 200\,GW is not reached.

\begin{figure}[tb]
    \centering
    \includegraphics[width=0.8\textwidth]{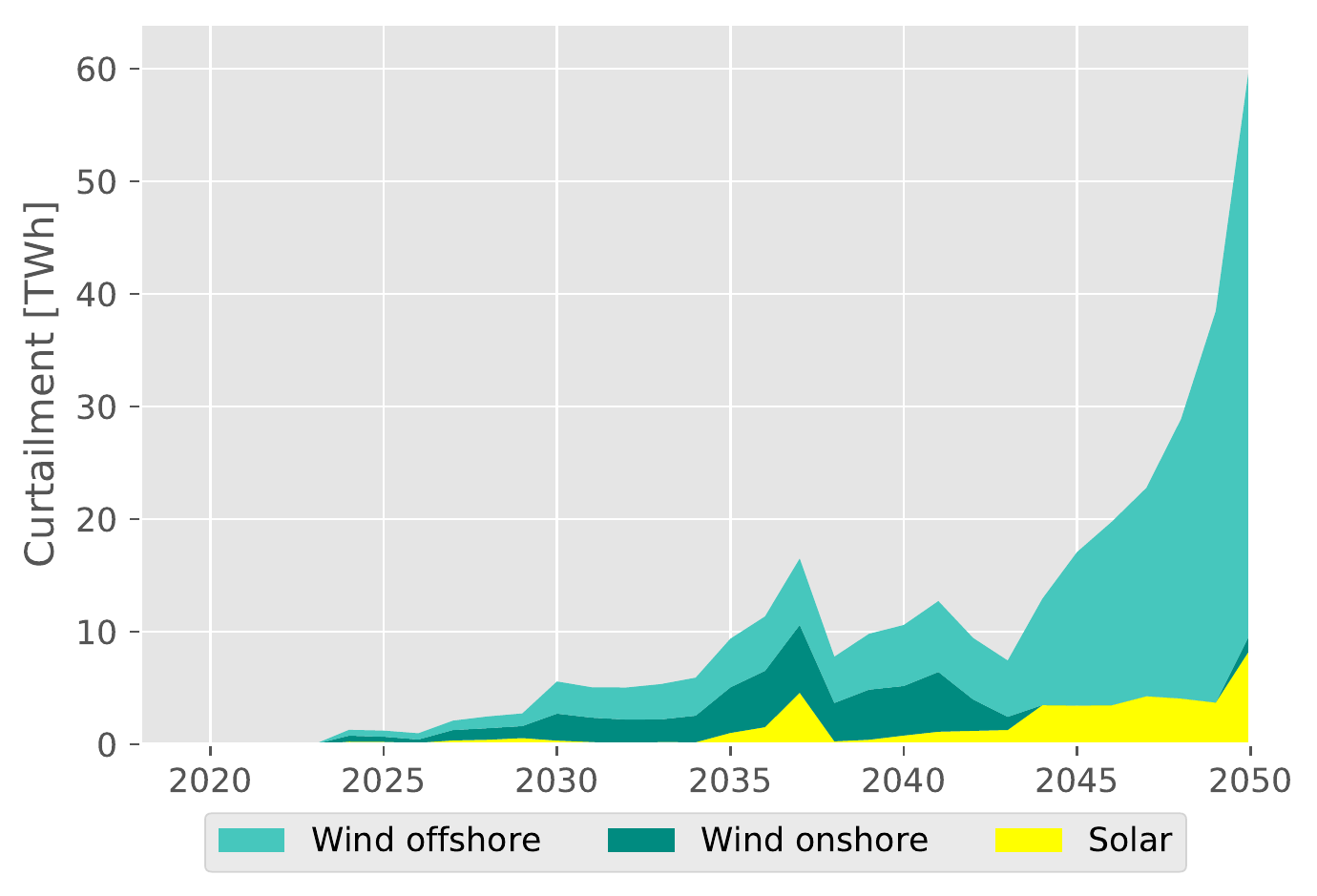}
    \caption{Electricity curtailment from non-dispatchable renewable energy technologies. Curtailment is calculated as the theoretical maximum amount of produced electricity subtracted by the actual amount of produced electricity per time step. Due to the calculation method it is possible to distinguish between the technologies. Curtailment is an important instrument to provide flexibility, however the amount of curtailed energy is potentially available for other processes (such as Power-to-Heat or variable industrial consumers).}
    \label{fig:curtailment}
\end{figure}

As the renewable capacities increase, curtailment becomes an essential element for the flexibilization of the power systems. In particular, we find that the amount of curtailment, shown in Fig.~\ref{fig:curtailment}, increases vastly by a factor of six from 2040 to 2050. Fig.~\ref{fig:curtailment} shows the development of curtailment by technology. We have calculated the curtailed amount of electricity E of each technology t with:
\begin{equation}
\mathrm{
    E_{curtailed,t} = E_{potential,t} - E_{actual,t}.
}
\end{equation}
The differentiation between the technologies are partly random, as in some time steps all technologies may lead to the same costs. Thus, the solver chooses randomly which technology is used. In reality, curtailment decisions are highly dependent on local factors, grid stability, transport distances and last but not least market mechanisms.

The reconversion of SNG to electricity provides another source of flexibility, which is established in the final years of the modeling period, in particular in 2050. The reconversion takes place in the conventional gas fired power plants (cf.~section~\ref{sec:temporal-patterns-and-resolution}), without contributing to the net CO$_2$ emissions. In this scenario the reconversion appears to be low (approx. 1\,TWh$_{el}$), however this might change with different scenario demands (cf.~Fig.~\ref{fig:03demand}) or if the emission cap is lowered further after 2050.

A remarkable development is observed in the operation of PtM plants: They change from flexibility consumers to flexibility providers. Before 2040 they were mostly operating at full capacity due to the high investment costs -- thus requiring other technologies to balance the variability of renewable power generation. This is no longer possible such that PtM plants must operate flexibly, too. We discuss PtM operation in more detail in section~\ref{sec:ptgop}.

\subsection{PtM Operation}
\label{sec:ptgop}

PtM is a key element for sector coupling and can also provide flexibility to the system. Due to the huge storage capacities in the natural gas grid, there is no temporal demand profile for SNG production to be satisfied as described in section~\ref{sec:temporal-patterns-and-resolution}. Rather, the SNG can be produced whenever it is most cost efficient to do so, including the option to build excess capacities. Thus, the PtM technology will be operated in a flexible way if it is beneficial for the system regarding total discounted system costs (cf.~Eq.~\ref{eq:objective_function}). In particular, investment costs for PtM are rather high such that a flexible operation with lower utilization rates is costly. Indeed, we observe a fundamental change in the operation of PtM. Initially, it runs at utilization rates close to 100\,\% and thus effectively consumes flexibility. Later, it is a key instrument to provide flexibility.

Fig.~\ref{fig:cap_vs_fng_ptm} shows the change of installed capacity (a) and system operation in terms of the average capacity factor (b) for PtM. Capacity utilization is well above 90\,\% until 2033. A first dip is observed around 2035, in the final years of the coal phase-out, but the utilization settles back to high values of approximately 80\,\% around 2040. A drastic change is observed in the last decade: PtM becomes a flexible consumer and acts as a demand side management device. The capacity utilization decreases steadily, until it is below 40\,\% in 2050. This decrease must be compensated by a strong increase in the capacity of PtM plants to meet the growing demand for SNG. That is, a massive excess capacity of PtM is required to meet the demand for flexibility in future renewable energy systems. This development is further promoted by the assumed cost degression for PtM plants (cf.~Table~\ref{tab:techno-economic-parameters}).

\begin{figure}[tb]
    \centering
    \includegraphics[width=0.8\textwidth]{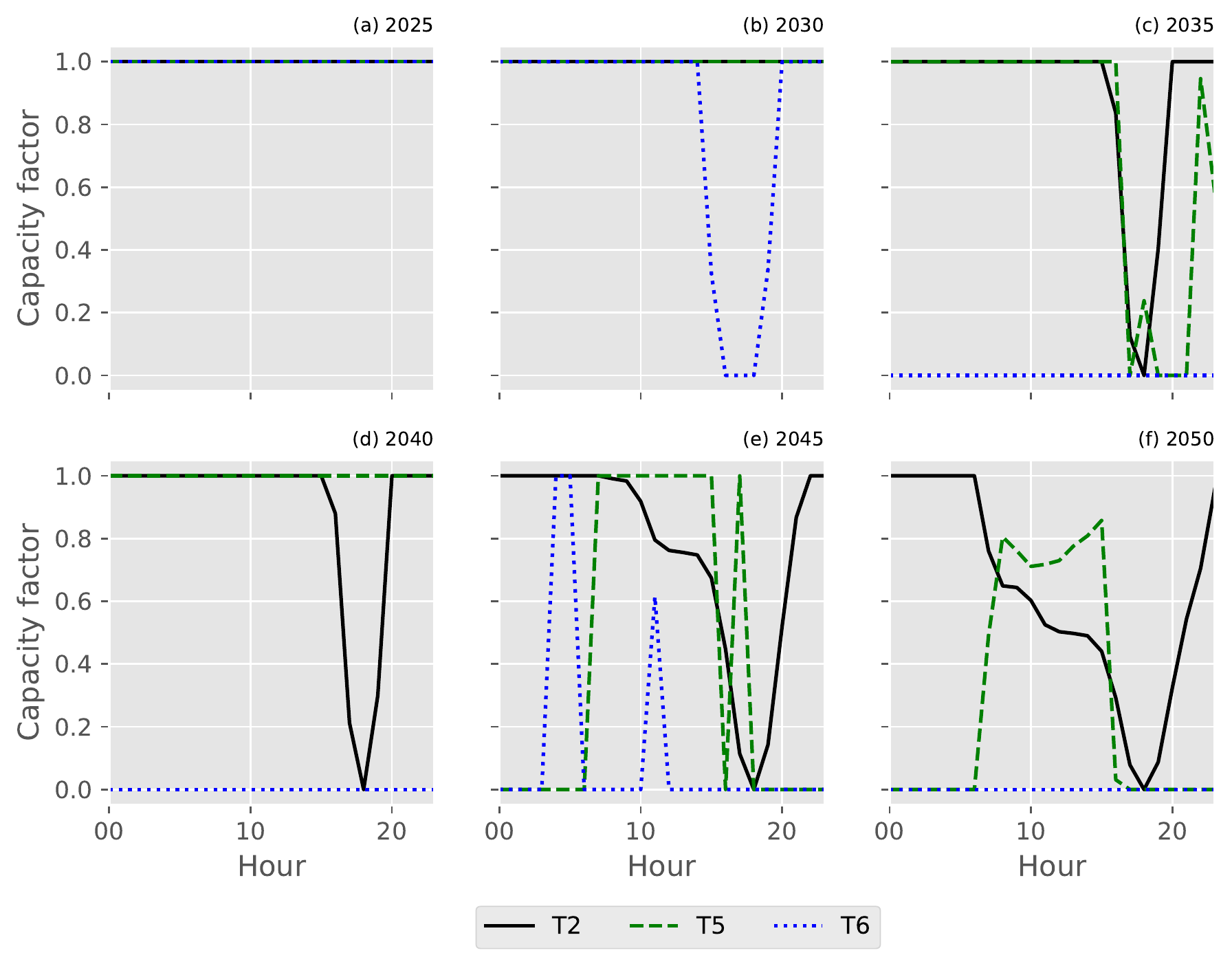}
    \caption{Load behavior of PtM plants for three type days and six years. T2 is characterized by low PV and high wind pentration, T5 shows the opposite case and T6 is represents a moderate penetration (cf.~Fig.~\ref{fig:typedays4x6} for detailed type day characterization). Already at the beginning of the modeled period the PtM plant operates partly flexible, e.g. decreasing production in the evening load peak. Later, it contributes significantly to a fully flexible system operation.}
    \label{fig:ptm_cfs_typedays}
\end{figure}

The fundamental change in system operation becomes most obvious in the development of the daily operation pattern, as shown for three of the twelve type days in Fig.~\ref{fig:ptm_cfs_typedays}. Each type day is different in resulting load and renewable generation (cf.~Fig.~\ref{fig:typedays4x6}), to which the PtM plant adapts its operation. In 2025, the PtM plant runs continuously at full capacity during all three type days. No adaption to the load is observed. In 2030 and 2035 this behavior is changed slightly for T5 and T6. PtM plants reduce their load in the evening, when the demand for electricity peaks due to the charging of BEVs while PV generation drops.
From 2040 onwards, the PtM load adapts to the systems requirements in a highly flexible way: From shutting down production for an entire day (T6, 2040) to alternately complete and zero capacity utilization (T5, 2045). We find that PtM performs crucial tasks for the entire energy system requiring a low average capacity utilization. In contrast, investors typically aim for a high capacity utilization to maximize revenues, in particular for technologies with high investment costs such as PtM. This dilemma emphasizes the challenges for market design and regulation in future energy systems: How to compensate system-friendly operation and how incentivize investments in low-utilization technologies?

Due to the gas grid storage assumptions SNG can be pre-produced and shifted to later periods, for storage purposes or to avoid transmission bottlenecks. This actually happens a few times (2020, 2026, 2038) in small amounts (below 1\,TWh$_{ch}$ each). Noteworthy, in 2040 and 2041 more than 6\,TWh$_{ch}$ is shifted to be released in 2042 and 2043. The influence of period shifting is likely to be negligibly small due to shifted amounts. However, this feature could play an important role in different demand scenarios for SNG.

\subsection{Residual load}
\label{sec:residualload}

Flexibility is needed to adapt to the variability of renewable power generation and the demand. The effective variability is conveniently analyzed in terms of the residual load, which we define as the total electricity generation (which has to equal the load) minus the generation from renewable sources. A classification of generation technologies as renewable/non-renewable is included in Table~\ref{tab:techno-economic-parameters}. Gas fired power plants are counted as renewable in their second mode of operation, i.e.~when they burn SNG instead of FNG, cf.~section~\ref{sec:temporal-patterns-and-resolution}. 
We note that the term residual load is used in different ways in the literature \cite{energypediainfo_residual}. We focus on the renewable aspect and do not distinguish between dispatchable and non-dispatchable technologies (which is another approach). Thus, the residual load and its development over time provides details how much fossil capacity is still required for how many hours during the year.

\begin{figure}[tb]
    \centering
    \includegraphics[width=0.7\textwidth]{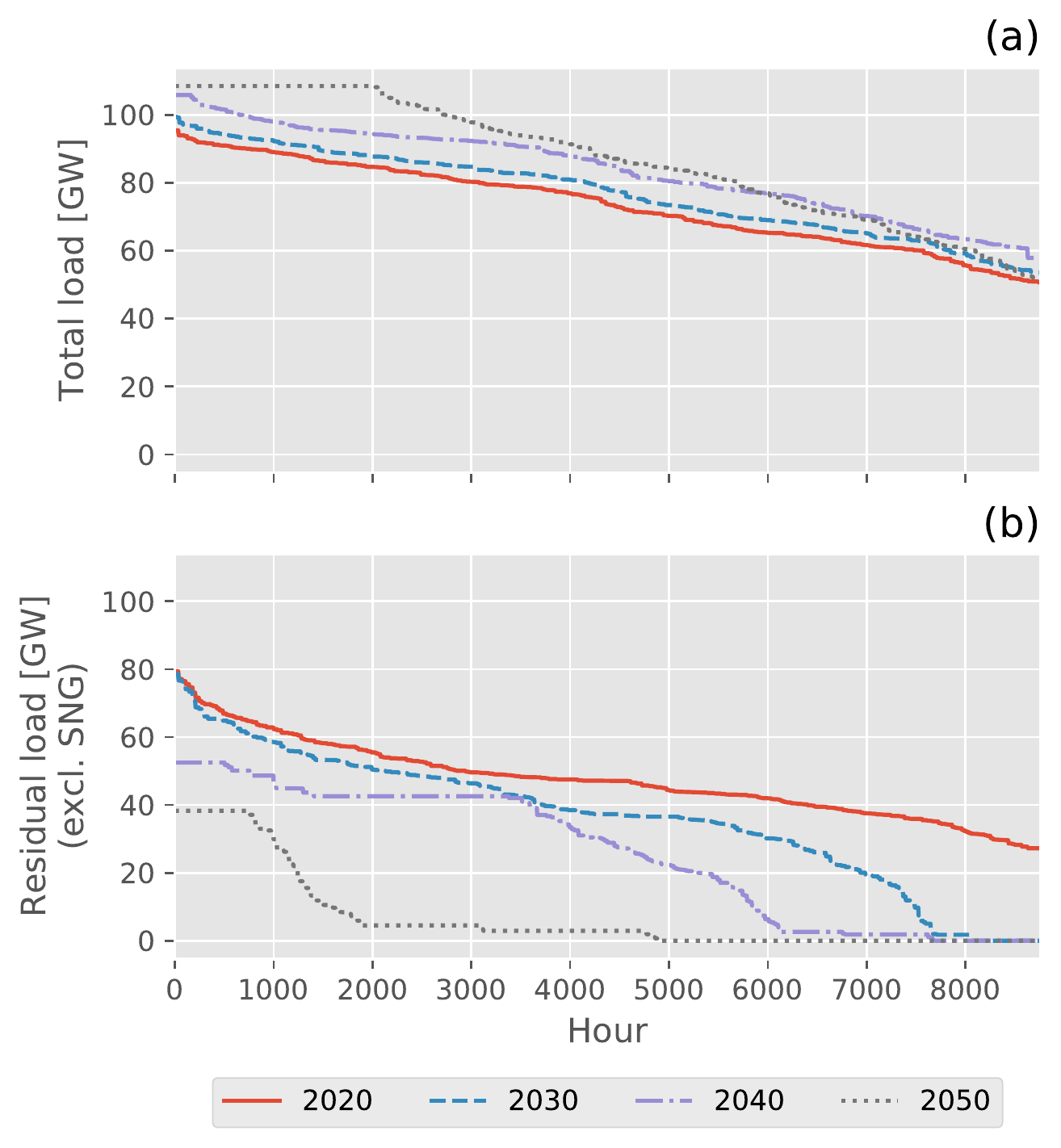}
    \caption{Annual load duration curve and resulting annual residual load duration curve. The annual load duration curve shows the lengths of load. In 2050 the load is highest because of additional demand (cf.~Fig.~\ref{fig:03demand}) from SNG and electricity for electromobility. The residual load is the load that is supplied by conventional technologies, such as gas fired power plants. Whenever synthetic natural gas is used as fuel in a gas fire power plant, it does not count as a conventional technology. Thus, the residual load does not include reconversion of SNG to electricity.}
    \label{fig:residual_load}
\end{figure}

The statistics of both the residual and non-residual load are represented using ordered annual duration load curves in Fig.~\ref{fig:residual_load}. Ordered instead of consecutive specifically allows to determine the load height and its total annual duration for the selected years. Furthermore, the area under the respective curves is the corresponding amount of electric energy.

The magnitude of the annual load duration curve (Fig.~\ref{fig:residual_load}~(a)) increases slightly with increasing years. This is due to the additional (and rising) demand of SNG and electricity for electromobility (cf.~Fig.~\ref{fig:03demand}). To satisfy resulting peaks (cf.~also~Fig.~\ref{fig:typedays4x6}) the maximum load is above 105\,GW in 2050 for more than 2000 hours during the year, which is an annual share of more than 20\,\%. This corresponds to an increase of peak load by approximately 20\,GW compared to 2018. Certainly, this represents a major challenge for the electricity grid, which must be expanded on many levels in the course of the transformation process \cite{gdp_tso_germany2019, bundesnetzagentur_grid}.

The development of the residual load (Fig.~\ref{fig:residual_load}~(b)) is dynamic. The different stages of the transformation process are clearly visible, which we have described in detail in section~\ref{sec:stages-of-transformation}. As more renewable energy is integrated into the energy system the demand for residual load changes. Until 2030 generation by conventional technologies still provide a large share of the electricity, as in providing large generation proportions over long periods. In 2030, short periods (<\,700\,h in total) with a vanishing residual are observed, i.e.~periods in which generation is completely based on renewables. The length of these periods increase to almost 4000 hours ($\approx$\,45\,\%) in 2050. However, even in 2050, roughly 1000 hours of very low renewable generation occur and have to covered by backup power plants. During these periods, the residual load is still above 30\,GW which is almost completely supplied by gas fired power plants. Hence, large capacities of dispatchable gas fired power plants are still required for backup tasks even in a 95\,\% CO$_2$ emission reduction scenario. These dispatchable backup plants are of major importance for system stability, yet average capacity utilization is rather low (cf.~Fig.~\ref{fig:cap_vs_fng_ptm}~(b)).

\section{Discussion}
\label{sec:conclusion}

Sector coupling is a key technology for the realization of a future sustainable energy system. In this article we have analyzed an efficient pathway for the transformation of the German energy system including the production of SNG and electricity for the use in the mobility and the industry sector. The pathway was derived from a techno-economic optimization model
with a high temporal and technological resolution implemented in the OSeMOSYS framework. Spatial aspects were not considered in this article.

Our result show that a reduction of GHG emissions by approximately 75\,\% with respect to the reference year 1990 is possible with no fundamental system change -- even if the additional demand for electricity for BEVs and the production of SNG is included. In our model, this reduction is achieved until 2040 mainly by three central measures: (1) a strong extension of renewable power generation, in particular solar PV, (2) the replacement of coal-fired power plants by gas-fired plants and (3) the expansion of battery storage capacities. All measures can be implemented with mature technologies. 

Our model takes into account the recent rapid progress in the development of solar PV, assuming significantly lower investment costs than many earlier studies. This rapid decrease in costs significantly affects the transformation pathway. PV is preferred to wind power even in a country like German with limited solar resources. In the model, PV capacities almost triple until 2030 while almost no additional wind capacity is built. This remarkable development could mitigate one of the most urgent problems in the the German energy transformation: the lacking acceptance for new onshore wind turbines. 

The energy transformation faces fundamental challenges after 2040 as emission caps are further decreased. Previously, flexibility was mainly provided by backup power plants fueled by fossil natural gas. As this technology option becomes unavailable, all flexibility options must be drawn to guarantee a balanced energy system operation. This development becomes most apparent for PtM, which changes its role from a consumer to a provider of flexibility. The utilization rate of PtM is high until 2040 due to the high investment costs. Afterwards, it drops dramatically to below 40\,\% in 2050 as PtM must adapt to variable renewable generation and load. The operation of PtM is a prime example of interdependency of different sectors in future energy systems. A reduction of the load is needed in particular in the evening, when PV generation drops and the electricity demand peaks due to the charging of BEVs. PtM, as an  interface to the industry sector, operates in response to the electricity and mobility sector.     

The development of further PtM emphasize that the energy transformation is not just a technological challenge, but also a socio-economic one. System stability requires a flexible use of PtM at low average capacity utilization. In contrast, investors typically aim for a high capacity utilization to maximize revenues, in particular in view of the high investment costs. This dilemma emphasizes urgent questions in market design and regulation in future energy systems: How to compensate system-friendly operation and how to incentivize investments in low-utilization technologies?

In conclusion, our modeling results show that the largest technological challenges for German energy transformation are yet to come. The next steps can be achieved with mature technologies, and even the problem of a lacking acceptance of wind power may be mitigated by focussing on PV instead. In the long run, fundamental changes in regulation and market design may be needed.


\end{document}